# High-power mid-infrared frequency comb source based on a femtosecond Er:fiber oscillator

**Feng Zhu,[1*] Holger Hundertmark,[2] Alexandre A. Kolomenskii,[1] James Strohaber,[1] Ronald Holzwarth,[2,3] and Hans A. Schuessler[1,4]**

*[1] Department of Physics and Astronomy, Texas A&M University, College Station, TX 77843-4242, USA*
*[2]Menlo Systems GmbH, D-82152, Martinsried, Germany*
*[3]Max-Planck-Institut für Quantenoptik, 85748 Garching, Germany*
*[4]Science Department, Texas A&M University at Qatar, Doha 23874, Qatar*
**Corresponding author: zhuf@physics.tamu.edu*

We report on a high-power mid-infrared frequency comb source based on a femtosecond Er:fiber oscillator with a stabilized repetition rate at 250 MHz. The mid-infrared frequency comb is produced through difference frequency generation in a periodically poled MgO-doped lithium niobate crystal. The output power is about 120 mW with a pulse duration of about 80 fs, and spectrum coverage from 2.9 to 3.6 µm. The coherence properties of the produced high-power broadband mid-infrared frequency comb are maintained, which was verified by heterodyne measurements. As the first application, the spectrum of a ~200 ppm methane-air mixture in a short 20 cm glass cell at ambient atmospheric pressure and temperature was measured.

Currently there is a large demand for gas detection systems in the mid-infrared (MIR) in many areas of science and technology. For these applications high repetition rate femtosecond lasers and frequency combs are being developed actively due to fast data acquisition rates, high sensitivity, and multi-target detection properties inherent of broadband frequency comb spectroscopy [1, 2]. The 3 to 4 µm MIR range is of particular interest since it contains strong absorption features of the C-H stretching vibrational mode of methane ($\nu_3$ band) and many other more complex hydrocarbons. For a multitude of measuring tasks in the booming natural gas industry, agriculture, atmospheric and geosciences researches, methane detection with real time monitoring, and quantifying different hydrocarbon isotopes are important [3, 4]. In addition there is a growing interest to detect volatile organic compounds such as benzene, toluene and ethyl benzene, which are precursors of atmospheric nanoaerosols and can contribute to poor indoor air quality.

Different versions of frequency comb spectroscopies have been developed since the invention of frequency comb [2, 5-11]. Broadband MIR frequency combs provide useful light sources for many spectroscopic applications. In particular, several MIR sources using single pass difference frequency generation (DFG) have been developed and are attractive because of their relative simplicity and the benefit of passive carrier-envelope offset frequency stabilization [12-14]. If the pump and signal fields are phase coherent and originated from the same source, the generated idler field is carrier envelope phase slip free and requires only stabilization of the comb spacing, which is relatively easy to implement by stabilizing the source repetition rate. Hence it was shown to provide a frequency synthesizer in the MIR and can be used in frequency standard applications [15]. Several MIR sources based on DFG have been reported [12-14]; however in some cases involving Raman shifting the coherence was reduced and even lost [13]. In addition, the available power levels have been moderate with about 1.5 mW at 4.7 µm having been reached [14].

In this letter, we demonstrate a high-power MIR frequency comb source based on a femtosecond Er:fiber oscillator with a stabilized repetition rate at 250 MHz. Applying self-phase modulation to up-convert a portion of the output from the Er:fiber oscillator, the coherence properties of the produced high-power broadband MIR frequency comb are maintained, which was verified by heterodyne measurements. In addition, we measured the spectrum of a ~200 ppm methane-air mixture in a short 20 cm glass cell at ambient atmospheric pressure and temperature as the first application.

A simplified diagram of the MIR femtosecond laser is presented in Fig.1. Intense ultrashort pulses centered at ~1.05 µm and ~1.55 µm are generated from a 250 MHz mode-locked Er:fiber oscillator. Half of the output from the Er:fiber oscillator is amplified by the first Er doped fiber amplifier (EDFA) to about 450 mW. The amplified pulses are coupled into a highly nonlinear fiber (HNLF) to broaden the spectrum and generate the required ~1.05 μm seed pulses. Due to the process of self-phase modulation, the comb characteristics of the original radiation at ~1.55 μm are transferred to the seed at ~1.05μm. After filtering, the ~1.05 μm seed pulse is amplified by a high-power Yb doped fiber amplifier (YDFA) to about 1.2 W. After an isolator and a compressor, the pulse duration of pump pulses centered at ~1.05 μm is about 90 fs. The other half of the output from the Er:fiber oscillator is amplified by the second EDFA to about 450 mW to generate single pulses centered at ~1.55 μm; the pulse duration is about 60 fs. The DFG occurs in a periodically poled MgO-doped lithium niobate (MgO:PPLN) crystal, where the pump pulse and signal pulse overlap in time and space, resulting in an idler wave in the MIR ($\lambda_i^{-1} = \lambda_p^{-1} - \lambda_s^{-1}$). Pump and signal pulses are combined at a dichroic mirror before being collinearly focused into the MgO:PPLN crystal. Temporal overlap of the pulses is achieved by introducing optical delay in the path of the signal pulse. Since both pump and signal pulses are not completely converted, a Ge long-pass filter (>2.4 μm) is used to select the idler pulses in the MIR.

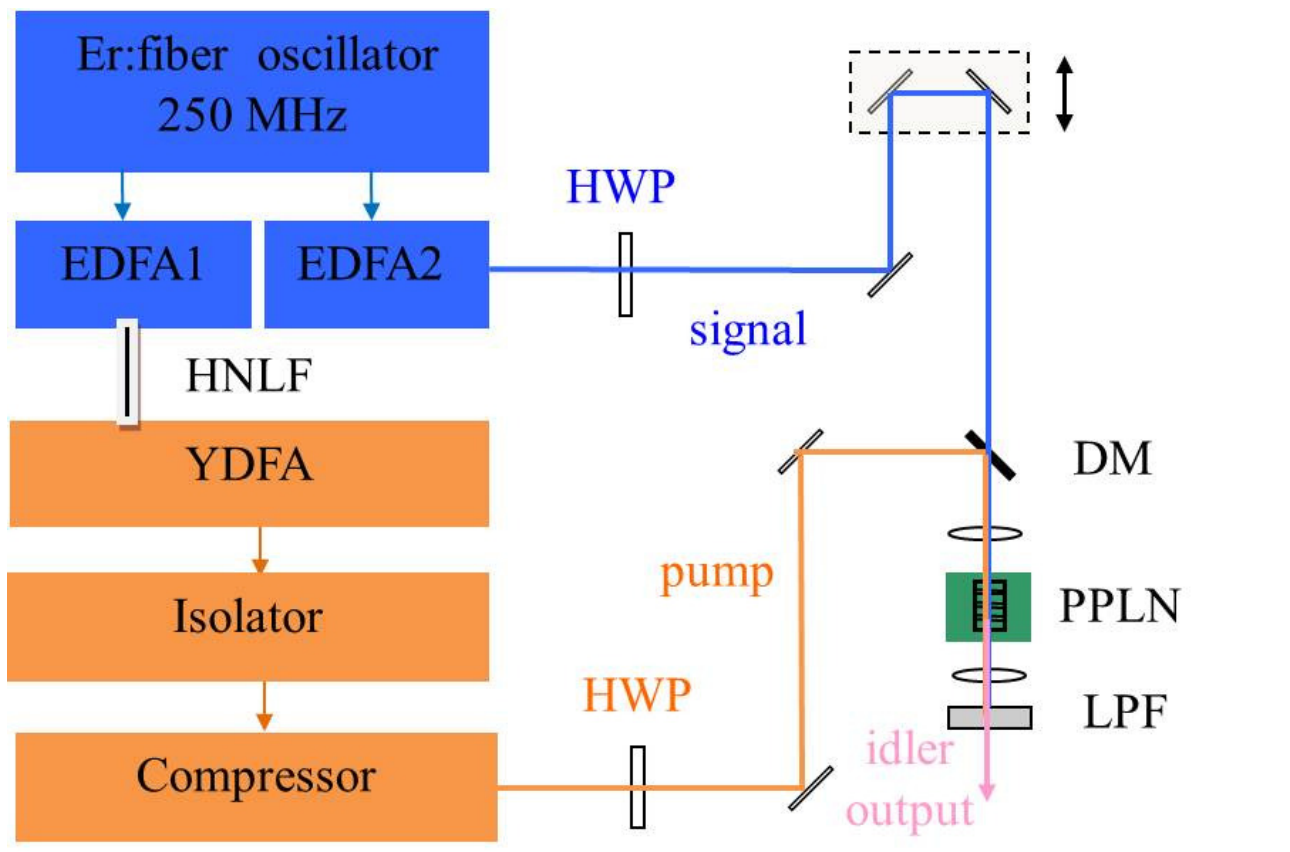

Fig. 1. (Color online) Simplified diagram of MIR femtosecond laser. EDFA, Er doped fiber amplifier; YDFA, Yb doped fiber amplifer; HNLF, highly nonlinear fiber; HWP, half waveplate; DM, dichroic mirror; PPLN, periodically poled lithium niobate; LPF, longpass filter.

To characterize the MIR femtosecond laser, we measured the spectrum, recorded the interferometric autocorrelation trace and verified the coherence of the MIR comb with heterodyne beat experiments.

We measured the spectrum of the MIR comb presented in Fig. 2 with a 0.3 meter McPherson 218 scanning monochromator and a highly sensitive thermal powermeter (Ophir, 3A). The absorption features in the spectrum are due to water vapor in the laboratory environment (~20% relative humidity, ~0.6% concentration in air, ~6 meters path length). By tuning the temperature of the MgO:PPLN crystal, the output spectrum and power can be slightly changed through different quasi-phase matching conditions. Experimentally, we determine the temperature of ~90 $^0$C for ~120 mW maximum output power in the spectral range from 2.9 to 3.6 μm showing a strong peak at the methane absorption band ~ 3.3 μm.

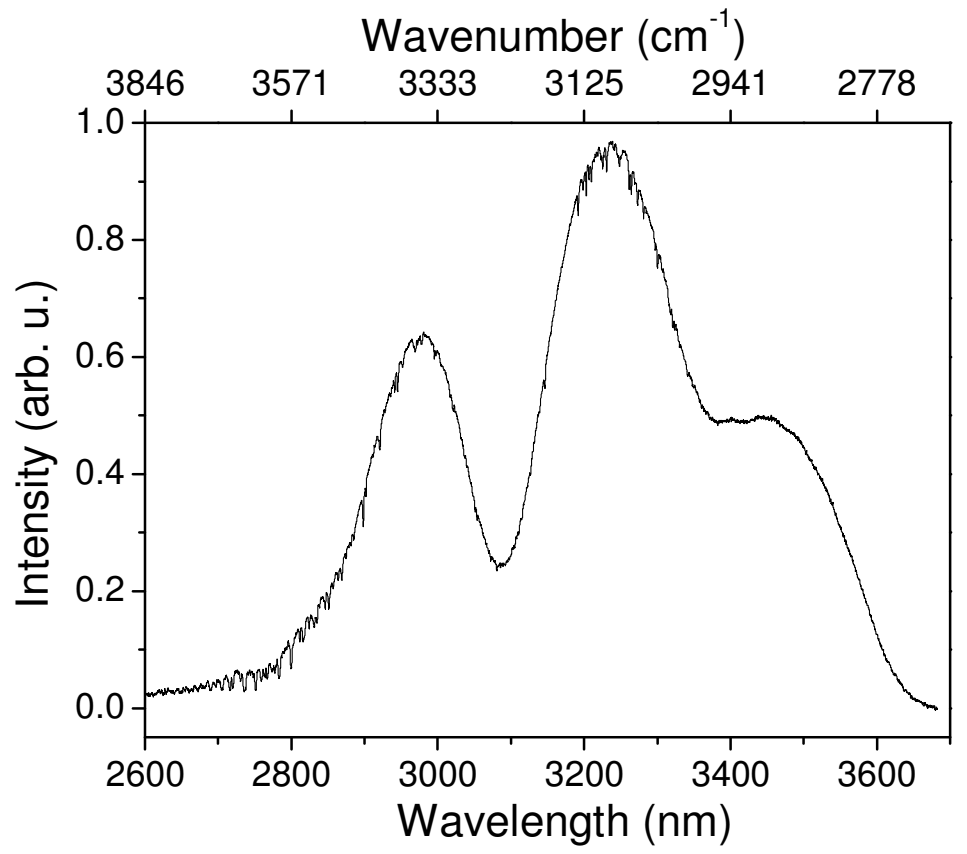

Fig. 2. Spectrum of the MIR femtosecond laser. The absorption features in the spectrum are due to water vapor in the laboratory environment.

The temporal characteristics of the MIR pulse were measured with a Michelson type interferometric autocorrelator using two-photon absorption with an InGaAs photodiode (Hamamatsu, G8376-03). The interferometric autocorrelation trace is presented in Fig.3, and corresponds to a pulse duration of about 80 fs assuming a Gaussian pulse shape.

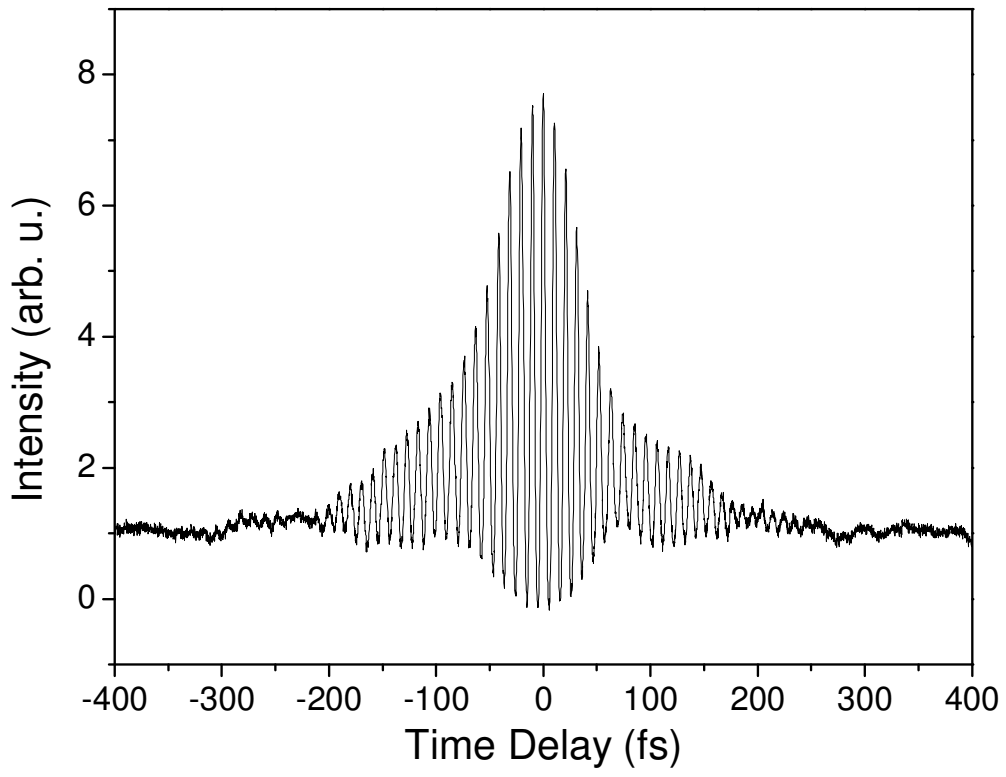

Fig. 3. Interferometric autocorrelation trace of the MIR femtosecond laser. The autocorrelation trace corresponds to a pulse duration of about 80 fs assuming a Gaussian pulse shape.

To verify the coherence we converted the MIR comb to a near infrared comb by second harmonic generation (SHG) with a $AgGaS_2$ crystal, because a fast MIR photo detector with required bandwidth, an accessible narrow-linewidth MIR continuous wave (CW) laser source as well as polarization optics for ~3.3 μm are lacking. About 500 μW of SHG radiation centered at ~1.6 μm was generated by focusing the MIR comb into a 1 mm thick $AgGaS_2$ crystal with type I phase matching (Eksma Optics, AGS-401H). The heterodyne beat note between the SHG comb and a 1.602 μm CW distributed feedback diode laser (NEL, NLK1556STG) was detected with a fast InGaAs detector (Menlo Systems, FPD510-F) of DC-250 MHz bandwidth. The heterodyne beat note is presented in Fig. 4. The resolution bandwidth is 300 kHz, and the sweep time is 20 ms. The spread in the beat note frequency is largely due to non-stabilized frequency of the CW diode laser.

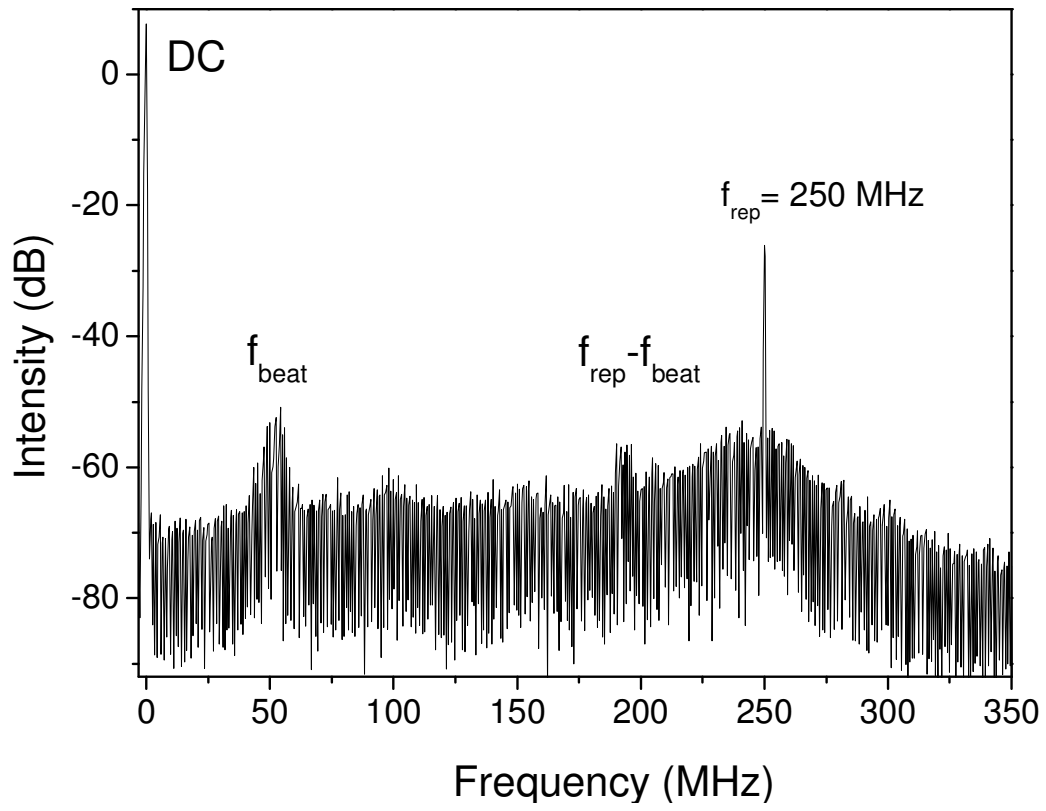

Fig. 4. Beat note between the CW laser and the near infrared frequency comb obtained by second harmonic generation from the MIR frequency comb. The resolution bandwidth is 300 kHz, and the sweep time is 20 ms. The spread in the beat note frequency is largely due to the non-stabilized frequency of the CW laser

As the first application of the MIR frequency comb, we measured the spectrum of ~200 ppm methane mixed with air in a short fused silica glass cell of 20 cm length at room temperature by inserting the sample gas cell in front of the scanning monochromator. The absorption spectrum is presented in Fig. 5(a). Because fused silica has strong absorption above 3.5 µm, the spectrum in Fig. 5(a) shows strong absorption in this region compared with Fig. 3. The experimentally determined transmittance is depicted in Fig. 5(b), and its comparison with calculations from the HITRAN data base [16] shows good agreement between measurement and literature values.

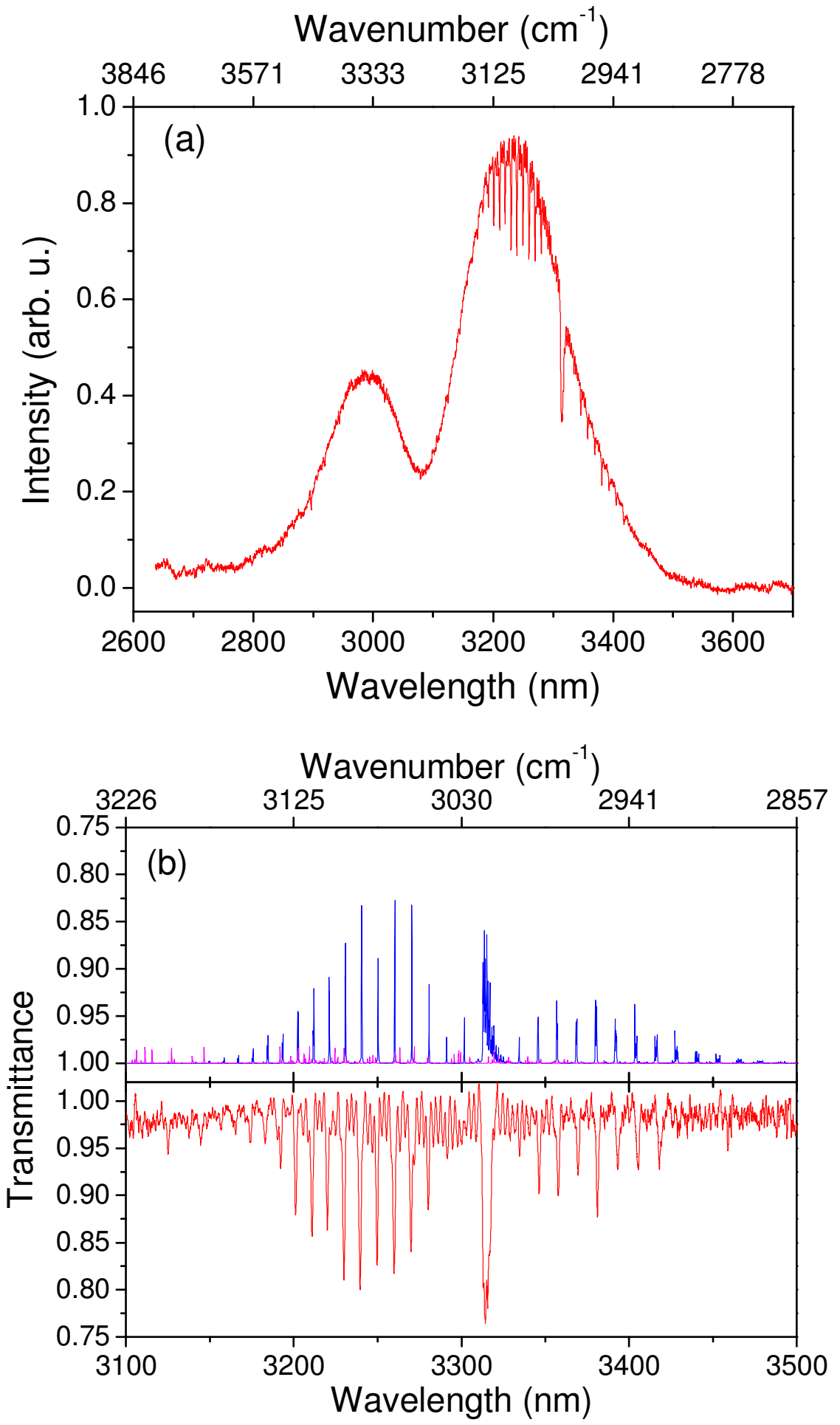

Fig. 5. (Color online) Methane and water vapor absorption spectrum (a) and experimentally determined transmittance compared with calculations from HITRAN database (b). The HITRAN calculation is inverted for clarity. The calculations are done for the experimental conditions: blue, 200 ppm methane in a 20 cm glass cell; purple, 0.6% water vapor in air, 6 m path length.

In summary, we demonstrated a high-power MIR frequency comb source based on a femtosecond Er:fiber oscillator with a stabilized repetition rate at 250 MHz. We substantiated the coherence of the MIR frequency comb and verified its suitability for spectroscopic applications by a straight-forward experiment of detecting methane in a short glass cell. The measured fundamental vibrational mode of methane at ~3.3 µm has a much larger absorption strength (~150:1) than the overtone in the near infrared range at ~1.65 µm [16], in which several parts per billion detection sensitivity has been demonstrated with a near infrared frequency comb [7]. With the developed MIR frequency comb source, and signal improvement techniques such as multipass [9] or cavity enhancement [7, 8], as well as the development of a fast and sensitive MIR detector, several parts per trillion level of the detection sensitivity is possible. Many techniques can benefit from the developed MIR frequency comb source, such as frequency metrology and frequency standard [15], dual frequency comb spectroscopy [5, 8, 9], upconversion spectroscopy [10, 11], and provide an ultrafast, highly sensitive platform for the gas detection and its applications.

This work is funded by the Qatar Foundation under the grant NPRP 09-585-1-087, the Robert A. Welch Foundation grant No. A1546, and the NSF grant No. 1058510.

**References**

1. Th. Udem, R. Holzwarth, and T.W. Hänsch, Nature **416**, 233 (2002).
2. A. Schliesser, N. Picque, and T. W. Hänsch, Nature Photonics **6**, 440 (2012).
3. F. Rust, Science **211**, 1044 (1981).
4. D. M. Jones, I. M. Head, N. D. Gray, J. J. Adams, A. K. Rowan, C. M. Rowan, C. M. Aitken, B. Bennett, H. Huang, A. Brown, B. F. J. Bowler, T. Oldenburg, M. Erdmann, and S. R. Larter, Nature **451**, 177 (2008).
5. F. Keilmann, C. Gohle, and R. Holzwarth, Opt. Lett. **29,** 1542 (2004).
6. S. A. Diddams, L. Hollberg, and V. Mbele, Nature **445**, 627 (2007).
7. M. J. Thorpe, and J. Ye, Appl. Phys. B **91**, 397 (2008).
8. B. Bernhardt, A. Ozawa, P. Jacquet, M. Jacqey, Y. Kobayashi, Th. Udem, R. Holzwarth, G. Guelachvili, T. W. Hänsch, and N. Picque, Nature Photonics **4**, 55 (2010).
9. A. M. Zolot, F. R. Giorgettta, E. Baumann, J. W. Nicholson, W.C. Swann, I. Coddington, and N. R. Newbury, Opt. Lett. **37**, 638 (2012).
10. T. W. Neely, L. Nugent-Glandorf, F. Adler, and S. A. Diddams, Opt. Lett. **37**, 4332 (2012).
11. J. S. Dam, P. Tidemand-Lichtenberg, and C. Pedersen, Nature Photonics, **6**, 788 (2012).
12. D. G. Winters, P. Schlup, and R. A. Bartels, Opt. Lett. **35**, 2179 (2010).
13. T. W. Neely, T. A. Johnson, and S. A. Diddams, Opt. Lett. **36**, 4020 (2011).
14. A. Ruehl, A. Gambetta, I. Hartel, M. E. Fermann, K. S. E. Eikema, and M. Marangoni, Opt. Lett. **37**, 2232 (2012).
15. M. A. Gubin, A. N. Kireev, A. V. Konyashchenko, P. G. Kryukov, A. S. Shelkovnikov, A. V. Tausenev, and D. A. Tyurikov, Appl. Phys. B **95**, 661 (2009).
16. HITRAN database, http://www.cfa.harvard.edu/hitran/